\documentclass[conference]{IEEEtran}
\usepackage{amsmath}
\usepackage{amssymb}
\usepackage{amsfonts}
\usepackage{graphicx}
\usepackage{float}
\usepackage[table, dvipsnames]{xcolor}
\usepackage{adjustbox}
\usepackage{booktabs}
\usepackage[short]{optidef}
\usepackage{multirow}
\usepackage{algorithm}
\usepackage{algpseudocode}
\usepackage{caption}
\usepackage{subcaption}
\usepackage{tabularray}
\usepackage{arydshln}
\usepackage{tikz}
\usepackage{multirow}
\usepackage{makecell}
\usepackage{booktabs}
\usepackage{color,soul}

\begin{document}

\title{Lightweight CNN-BiLSTM based Intrusion Detection Systems for Resource-Constrained IoT Devices}

\author{\IEEEauthorblockN{Mohammed Jouhari\IEEEauthorrefmark{1}, Mohsen Guizani\IEEEauthorrefmark{2}
}
\IEEEauthorblockA{\IEEEauthorrefmark{1}LSIA Laboratory, Moroccan School of Engineering Sciences (EMSI), Tanger, Morocco\\
\IEEEauthorrefmark{2}Machine Learning Department,
Mohamed Bin Zayed University of Artificial Intelligence (MBZUAI),
Abu Dhabi, UAE\\
m.jouhari@emsi.ma, mohsen.guizani@mbzuai.ac.ae}
} 

\maketitle

\begin{abstract}
Intrusion Detection Systems (IDSs) have played a significant role in detecting and preventing cyber-attacks within traditional computing systems. It is not surprising that the same technology is being applied to secure Internet of Things (IoT) networks from cyber threats. The limited computational resources available on IoT devices make it challenging to deploy conventional computing-based IDSs. The IDSs designed for IoT environments must also demonstrate high classification performance, utilize low-complexity models, and be of a small size. Despite significant progress in IoT-based intrusion detection, developing models that both achieve high classification performance and maintain reduced complexity remains challenging. In this study, we propose a hybrid CNN architecture composed of a lightweight CNN and bidirectional LSTM (BiLSTM) to enhance the performance of IDS on the UNSW-NB15 dataset. The proposed model is specifically designed to run onboard resource-constrained IoT devices and meet their computation capability requirements. Despite the complexity of designing a model that fits the requirements of IoT devices and achieves higher accuracy, our proposed model outperforms the existing research efforts in the literature by achieving an accuracy of 97.28\% for binary classification and 96.91\% for multiclassification.
\end{abstract}

\section{Introduction}
The infrastructure of smart cities is increasingly vulnerable to cyberattacks as they expand and integrate ubiquitous IoT networks \cite{10122600}. These attacks can pose a threat to public safety by disrupting critical services and compromising sensitive data. In interconnected networks like these, intrusion detection systems (IDS) play a crucial role in real-time monitoring and identification of malicious activity before it causes significant damage \cite{10379640}. Although existing IDS solutions have shown promising results, they often face challenges in smart city environments. Traditional rule-based systems struggle to adapt to constantly evolving attack patterns, while complex deep learning models can be computationally expensive and resource-intensive for resource-constrained smart city devices \cite{10279198}. These IoT devices, which include temperature sensors and smart devices for monitoring light bulbs or thermostats, are characterized by their small size and limited functionality. The data collected by these devices is transmitted via the internet for storage and analysis, making security measures necessary to protect personal, private, or sensitive information \cite{s19081935}.

IoT has enhanced individuals' lifestyles by introducing automated services. As a result of this unchecked growth, privacy and security concerns have become increasingly important. Addressing these concerns is a challenging task due to the inherent complexity of dynamic and diverse IoT ecosystems. Consider protecting such a dynamic environment as a smart city or smart home which are characterized by their constrained visibility of the network, and the lack of professional support for the devices after their first use. This makes the traditional network security methods ineffective due to the dynamic nature of this environment. The risk to user privacy and the environment's general security is further increased by the fact that many interconnected IoT devices are poorly designed and do not receive regular security updates. Thus, to fully realize the potential of IoT-based smart environments, creative solutions that prioritize security and privacy while also being able to adapt to this ever-changing environment are required \cite{9528421}.

Previous research on smart city IDS has predominantly utilized traditional machine learning (ML) techniques like decision trees and Support Vector Machines (SVMs) to enhance IDS performance. While these methods deliver respectable accuracy, they require extensive manual feature engineering and often suffer from high rates of false positives. In contrast, deep learning methods, especially convolutional neural networks (CNNs), have emerged as promising alternatives due to their ability to autonomously extract features from network traffic data. However, their substantial computational demands render them impractical for smart city devices with limited resources. In this study, we introduce several key contributions:

\begin{enumerate}
    \item We propose a novel hybrid approach that integrates lightweight CNNs with bidirectional Long Short-Term Memory (LSTM) networks to achieve efficient and accurate intrusion detection in IoT networks.
    \item We develop a customized CNN architecture optimized for IoT devices, featuring fewer layers and parameters to facilitate efficient data processing. This architecture efficiently extracts critical features from network traffic, which are subsequently processed by a bidirectional LSTM for accurate classification of normal and malicious activities.
    \item Our hybrid model combines the strengths of both CNNs and LSTMs, addressing the limitations inherent in each method.
    \item To tackle the challenge of imbalanced datasets, we implement a weighted loss function in our multiclassification model. The weights are assigned based on the relative frequency of each attack type within the UNSW-NB15 dataset, enhancing model fairness and accuracy.
\end{enumerate}

The remainder of this paper is structured as follows: Section 2 reviews relevant literature related to our study. Section 3 offers a detailed description of our proposed hybrid intrusion detection approach, including the lightweight CNN-BiLSTM architecture, and outlines the dataset and evaluation metrics used in our analysis. Section 4 presents the experimental results and discusses our findings. Finally, Section 5 concludes the paper by summarizing our contributions and outlining future directions for this research.

\section{Related work}
The authors in \cite{Globecom2023hafsa} tackle anomaly detection in IoT networks by proposing a 1D CNN-LSTM model. This model leverages CNNs for spatial feature extraction and LSTMs for temporal analysis, achieving 99.20\% accuracy and 0.80\% false alarm rate on the Bot-IoT dataset. It offers both binary and multi-class detection, surpassing other methods and demonstrating potential for real-world applications. Our previous paper \cite{10000726} investigated a novel approach to improve anomaly detection in IoT networks using conditional Generative Adversarial Networks (cGANs). We addressed the issue of unbalanced data and lack of specific attack samples by generating realistic adversarial malware examples. This training process enhanced our CNNLSTM-based IDS model, increasing its accuracy in detecting theft attacks by 40\%. 

Many researchers have explored how to improve IDS performance using the UNSW-NB15 dataset, which is popular for its realistic and diverse network traffic data, including both normal and attack activities. A Feed-Forward Neural Network and the UNSW-NB15 dataset are used in the paper \cite{8859773} to investigate the use of Deep Learning for Network IDS. In the article, it is argued that traditional machine learning fails to detect intrusions accurately because of the complexity of modern network threats. As a result of their experiments, the authors were able to achieve high accuracy and lower false alarm rates compared to existing machine learning models by finding the optimal activation functions and features. 
In this paper \cite{Yin2023}, the authors proposed IGRF-RFE, a hybrid feature selection method for network intrusion detection that combined filter and wrapper techniques. IGRF-RFE successfully identified and eliminated irrelevant features, reducing the feature count from 42 to 23 while simultaneously improving anomaly detection accuracy from 82.25\% to 84.24\% on the UNSW-NB15 dataset. These results suggest that IGRF-RFE could have been a valuable tool for building more efficient and accurate IDSs.

In \cite{8983598}, the authors explored the use of Support Vector Machines (SVMs) with a new scaling method to detect network intrusions in the modern UNSW-NB15 dataset. They compared their method to previous approaches on both binary and multi-class classification tasks. The paper highlights the effectiveness of SVMs with a new scaling method for intrusion detection on the more realistic UNSW-NB15 dataset, especially compared to older methods and datasets. The authors in \cite{saurabh2022lbdmids} proposed a new Network Intrusion Detection System (NIDS) for protecting Internet of Things (IoT) devices from cyberattacks. The system used LSTM layer, which could learn from past data to identify both known and unknown attacks. This showed an improvement over older methods that struggled to detect new attacks. The researchers tested their system on two large datasets of network traffic and found it to be more accurate than previous methods, while also reducing false alarms. 

\section{Hybrid CNN MODEL FOR THE UNSW-NB15 DATASET}
\subsection{Description of the UNSW-NB15 dataset}
The UNSW-NB15 dataset provides a realistic snapshot of modern network traffic, encompassing both routine activity and a diverse range of nine contemporary attack types, including DoS, backdoors, and reconnaissance. Its clear organization with ten distinct data types and balanced representation of normal and malicious traffic makes it a valuable resource for developing and evaluating IDS. Refer to Table \ref{tab:unsw_categories} for detailed information on each data type.

\begin{table}[b!]
\centering
\renewcommand{\arraystretch}{1.3}
\caption{Categories of the UNSW-NB15 Dataset}
\label{tab:unsw_categories}
\begin{tabular}{clp{5.5cm}}
\toprule
\textbf{No.} & \textbf{Type} & \textbf{Description} \\ 
\bottomrule
1 & Normal & Routine, legitimate network traffic \\ \hline
2 & Analysis & Attempts to uncover vulnerabilities through web application attacks \\ \hline
3 & Backdoor & Covertly bypasses authentication for unauthorized remote access \\ \hline
4 & DoS & Disrupts system resources through memory-based attacks \\ \hline
5 & Exploits & Leverages network errors or bugs for malicious purposes \\ \hline
6 & Fuzzers & Crashes systems by overwhelming them with random data \\ \hline
7 & Generic & Exploits block-cipher vulnerabilities using hash functions \\ \hline
8 & Reconnaissance & Gathers information to evade security controls \\ \hline
9 & Shellcode & Executes code to exploit software vulnerabilities \\ \hline
10 & Worms & Self-replicating malware that spreads across systems \\ \bottomrule
\end{tabular}
\end{table}

This work utilizes the UNSW-NB15 dataset, which comprises 257,673 data instances (175,341 training and 82,332 testing), carefully split without redundancy to ensure reliable NIDS evaluations. Both sets share identical attack types, further detailed in Table \ref{tab:unsw_categories}. Each data point boasts 44 features categorized into six groups: flow, basic, content, time, additional generated (divided into general-purpose and connection features), and labeled attack classes. Features \#1-\#30 represent core network packet information, while additional and connection features (\#31-\#42) provide deeper insights. Labeled features (\#43-\#44) serve for supervised learning tasks (refer to Table \ref{tab:unsw_features}). This comprehensive dataset offers a robust platform for exploring intrusion detection in modern networks such as the IoT network in smart cities.

\begin{table}[t!]
\centering
\vspace{4mm}
\renewcommand{\arraystretch}{1.5}
\caption{Features of UNSW-NB 15 dataset}
\label{tab:unsw_features}
\begin{tabular}{clcl}
\toprule
\textbf{No.} & \textbf{Name of Feature} & \textbf{No.} & \textbf{Name of Feature} \\ \bottomrule
1 & proto & 23 & response\_body\_len \\ \hline
2 & rate & 24 & sinpkt \\ \hline
3 & dur & 25 & dinpkt \\ \hline
4 & service & 26 & sjit \\ \hline
5 & state & 27 & djit \\ \hline
6 & spkts & 28 & tcprtt \\ \hline
7 & dpkts & 29 & synack \\ \hline
8 & sbytes & 30 & ackdat \\ \hline
9 & dbytes & 31 & ct\_srv\_src \\ \hline
10 & sttl & 32 & ct\_srv\_dst \\ \hline
11 & dttl & 33 & ct\_src\_ltm \\ \hline
12 & sload & 34 & ct\_dst\_ltm \\ \hline
13 & dload & 35 & ct\_dst\_src\_ltm \\ \hline
14 & sloss & 36 & ct\_src\_dport\_ltm \\ \hline
15 & dloss & 37 & ct\_dst\_sport\_ltm \\ \hline
16 & swin & 38 & ct\_state\_ttl \\ \hline
17 & dwin & 39 & is\_ftp\_login \\ \hline
18 & stcpb & 40 & ct\_ftp\_cmd \\ \hline
19 & dtcpb & 41 & ct\_flw\_http\_mthd \\ \hline
20 & smeansz & 42 & is\_sm\_ips\_ports \\ \hline
21 & dmeansz & 43 & attack\_cat \\ \hline
22 & trans\_depth & 44 & label \\ \bottomrule
\end{tabular}
\end{table}

\subsection{Lightweight CNN-BiLSTM}
Traditional IDS often rely on rule-based systems or shallow machine learning models, which can struggle with the evolving nature of cyberattacks and the resource constraints of resource-constrained IoT devices. To address these challenges, we propose a novel hybrid model that combines the strengths of lightweight convolutional neural networks (CNNs) with bidirectional Long Short-Term Memory (BiLSTM) layers \cite{tatsunami2022sequencer} for effective intrusion detection in IoT networks. The first stage of our model utilizes a lightweight CNN architecture specifically designed for efficient processing on edge devices. This CNN employs fewer layers and parameters compared to standard CNNs, reducing computational complexity and memory footprint while maintaining acceptable accuracy. The convolutional layers extract spatial features from the network traffic data, identifying patterns and relationships between different data points. 

\begin{figure}[t!]
	\centering
 \vspace{3mm}
	\includegraphics[width=0.98\linewidth]{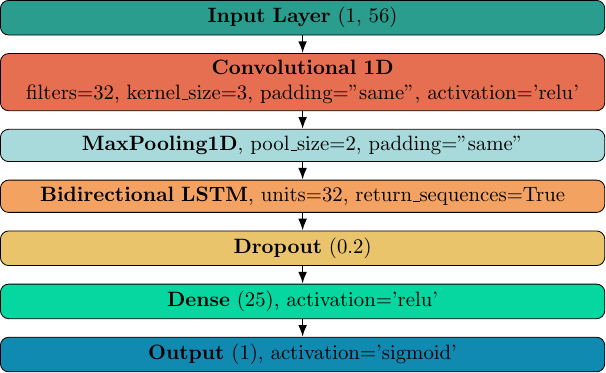}
	\caption{The architecture of our Lightweight CNN-BiLSTM Model with 7841 Trainable Parameters for UNSW-NB15 Dataset Classification}
	\label{combined}
 \vspace{-3mm}
\end{figure}

\begin{figure}[t!]
	\centering
 \vspace{4mm}
	\includegraphics[width=0.98\linewidth]{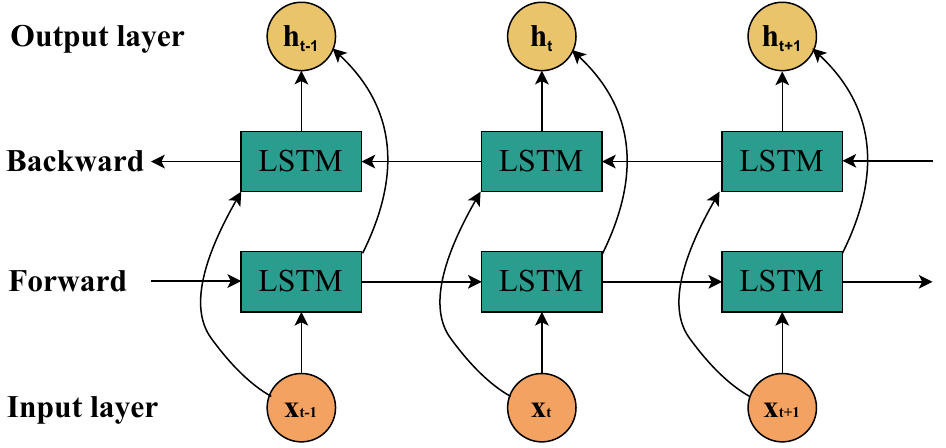}
	\caption{BiLSTM architecture.}
	\label{BiLSTM}
\end{figure}

The extracted features are then fed into a BiLSTM layer, which excels at capturing temporal dependencies within sequential data. Figure \ref{combined} present the architecture of our proposed CNN-BiLSTM model. This is particularly valuable for network traffic analysis, as cyberattacks often exhibit specific patterns over time. The BiLSTM analyzes the sequence of features, learning the underlying dynamics and identifying anomalies that deviate from normal network behavior. The outputs from the CNN and BiLSTM layers are combined, leveraging the strengths of both approaches. The combined features are then fed into a final decision-making layer, such as a sigmoid classifier, which determines the class of the network traffic (normal or malicious). The BiLSTM model is comprised of a pair of LSTM units. One LSTM unit processes the input sequence in the forward direction, while the other LSTM unit processes it in the reverse direction. In Figure \ref{BiLSTM}, we can see that the forward LSTM analyzes the input sequence from the start to the end, while the backward LSTM operates in the opposite direction. At each time step, the hidden states of the two LSTMs are combined or concatenated to create the final hidden state sequence. This combined sequence is then forwarded to the subsequent layer. The mathematical representation of this process is provided by Equations:

\begin{align}
\mathbf{h}_t^{(f)} &= \text{LSTM}^{(f)}\left(\mathbf{x}_t, \mathbf{h}_{t-1}^{(f)}, \mathbf{c}_{t-1}^{(f)}\right) \\
\mathbf{h}_t^{(b)} &= \text{LSTM}^{(b)}\left(\mathbf{x}_t, \mathbf{h}_{t+1}^{(b)}, \mathbf{c}_{t+1}^{(b)}\right) \\
\mathbf{h}_t &= [\mathbf{h}_t^{(f)}; \mathbf{h}_t^{(b)}]
\end{align}
where $\mathbf{h}_t^{(f)}$ and $\mathbf{h}_t^{(b)}$ represent the forward and backward hidden states at time step t, respectively. $\mathbf{c}_{t-1}^{(f)}$ and $\mathbf{c}_{t-1}^{(b)}$ represent the forward and backward cell states at time step t, respectively. $\mathbf{x}_t$ represents the input vector at time step t. $\mathbf{h}_t$ represented the concatenation of the forward and backward hidden states into a single vector.

\subsection{Measuring NIDS Performance: Key Metrics}
To assess the effectiveness of NIDS, we analyze key metrics derived from the confusion matrix, which maps the system's accuracy in classifying network traffic. We consider in the work the following parameters:

\begin{enumerate}
\item \textbf{Accuracy:} Measures the overall correctness of the NIDS, reflecting the proportion of correct predictions (both attacks and normal traffic) out of all predictions.
\begin{equation*}
    \text{Accuracy} = \frac{\text{( TP + TN )}}{\text{( TP + TN + FP + FN )}}
\end{equation*}
\item \textbf{Recall (also known as Detection Rate or Sensitivity):} Measures the NIDS's ability to correctly identify actual attacks. It focuses on how many of the actual attacks were correctly detected. 
\begin{equation*}
    \text{Recall} = \frac{\text{TP}}{\text{TP + FN}}
\end{equation*}
\item \textbf{Precision:} Measures the proportion of predicted attacks that were actually correct. It focuses on how many of the predicted attacks were true positives.
\begin{equation*}
    \text{Precision} = \frac{\text{TP}}{\text{TP + FP}} 
\end{equation*}
\item \textbf{F1-Score:} Combines precision and recall into a single metric, providing a balanced view of the NIDS's performance.
\begin{equation*}
    \text{F1-Score} = \frac{\text{2 * (Precision * Recall)}}{\text{Precision + Recall}}  
\end{equation*}
\end{enumerate}

While overall accuracy, encompassing both attack and normal traffic classification, plays a crucial role in assessing NIDS performance, its limitations within imbalanced datasets become readily apparent. In such scenarios, we prioritize recall, ensuring minimal missed attacks, as even a single undetected threat can have significant consequences. Conversely, high precision, minimizing false alarms, becomes vital to avoid resource drain and maintain operational efficiency. Ultimately, the F1-score, harmonizing these dual perspectives, offers a more nuanced understanding of the NIDS's efficacy, balancing attack detection with false alarm mitigation for robust network security.

\section{EXPERIMENTAL RESULTS AND ANALYSIS}
Our experiments, powered by Python and TensorFlow on a standard PC, investigated the feasibility of Lightweight CNNs for intrusion detection compared to traditional ML methods. Leveraging the versatile UNSW-NB15 dataset for multi-class classification, we built a custom CNN architecture within TensorFlow to assess its performance against established methods. This setup enabled a clear comparison of their effectiveness and resource demands, providing valuable insights for deploying CNNs in real-world network security scenarios.

\begin{figure}[t!]
	\centering
 \vspace{4mm}
	\includegraphics[width=0.88\linewidth]{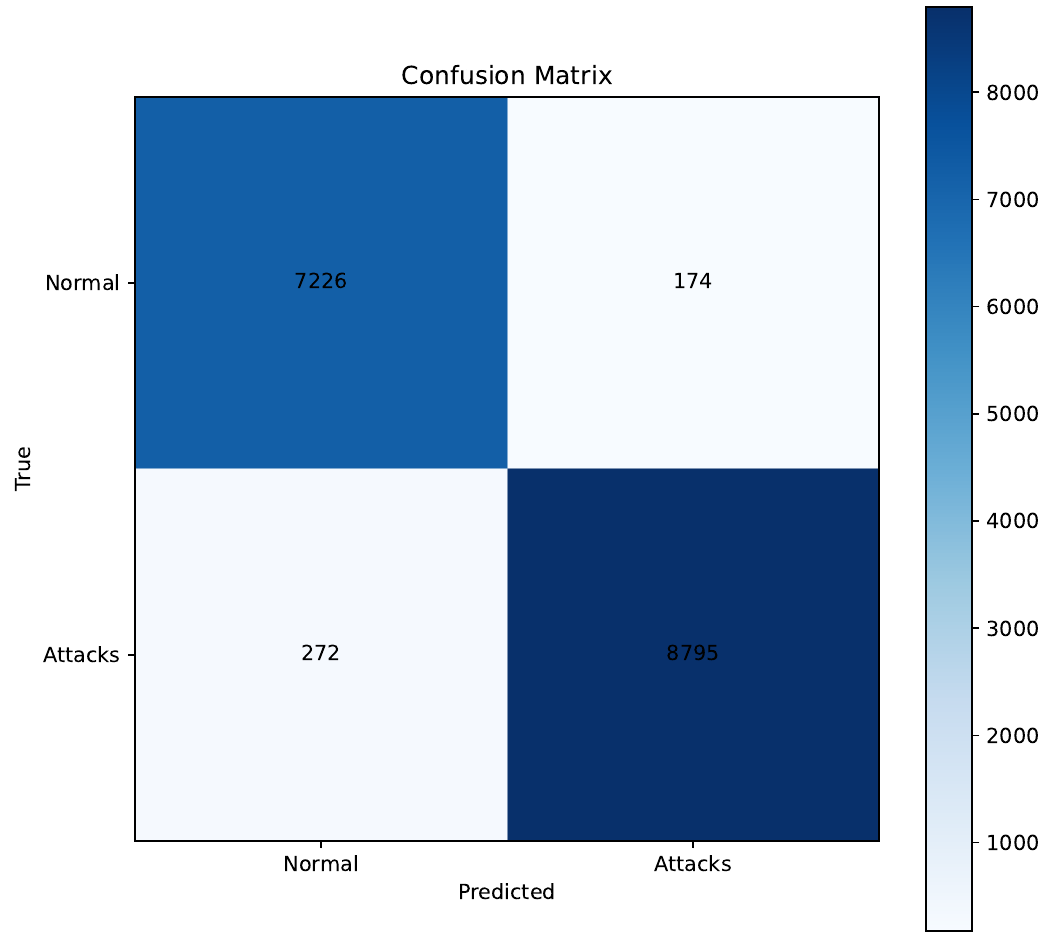}
	\caption{CNN-BiLSTM confusion matrix for binary classification of attacks.}
	\label{matrix}
 \vspace{-2mm}
\end{figure}

The confusion matrix in Figure \ref{matrix} presents the performance of our proposed model in classifying attacks on the UNSW-NB15 dataset. The horizontal axis (X-axis) represents the model's predicted attack categories for the testing data, while the vertical axis (Y-axis) shows the actual attack categories based on the ground truth labels. Our model achieved 7,226 true positives and 8,795 true negatives, correctly identifying both normal and attack instances. However, it also generated 272 false negatives (missed attacks) and 174 false positives (incorrectly identified attacks). While this translates to an overall accuracy of 97\%, as mentioned in the previous figure, we are committed to exploring methods to address the remaining false positives and negatives.

\begin{table*}[t!]
\centering
\vspace{4mm}
\renewcommand{\arraystretch}{1.3}
\caption{Performance Comparison of CNN-BiLSTM and Traditional ML Models for Binary Classification in UNSW-NB15}
\label{eval}
\begin{tabular}{l c c c c c c c }
\toprule
 & \textbf{Accuracy} & \textbf{Recall} & \textbf{Precision} & \textbf{F1-Score} & \textbf{time to train (s)} & \textbf{time to predict (s)} & \textbf{total time (s)} \\ \bottomrule
\textbf{Logistic Regression} & 92.80\% & 92.80\% & 92.83\% & 92.80\% & 1.2 & 0.0 & 1.2 \\ 
\textbf{KNN} & 95.04\% & 95.04\% & 95.09\% & 95.05\% & 0.0 & 15.3 & 15.3 \\ 
\textbf{Decision Tree} & 96.56\% & 96.56\% & 96.56\% & 96.56\% & 1.0 & 0.0 & 1.0 \\ 
\textbf{Gradient Boosting Classifier} & 95.85\% & 95.85\% & 95.86\% & 95.85\% & 27.4 & 0.0 & 27.4 \\ 
\textbf{MLP} & 96.34\% & 99.99\% & 55.05\% & 70.60\% & 21.3 & 21.9 & 43.2 \\  
\textbf{LSTM}  & 96.49\% & 96.49\% & 96.49\% & 96.49\% & 70.9 & 22.4 & 93.3 \\ 
\textbf{Proposed model} & 97.28\% & 96.44\% & 98.59\% & 97.43\% & 122.5 & 3.8 & 126.3 \\ \bottomrule
\end{tabular}
\end{table*}

Table \ref{eval} presents a performance evaluation of our CNN-BiLSTM model compared to traditional ML models from the literature. We assessed accuracy, recall, precision, and F1-score. While precision and recall are typically used for imbalanced data, we primarily use precision here to gauge our model's overfitting prevention capabilities. Combining these parameters through the F1-score provides a comprehensive understanding of our model's performance. Furthermore, we considered training and inference times to demonstrate the proposed model's suitability for resource-constrained IoT devices. As shown in the table, our model outperforms existing models in terms of accuracy, achieving 97.28\% compared to the best LSTMs using Keras (96.49\%). This suggests that our model can more effectively detect attacks on the UNSW-NB15 dataset. Our model also achieves the highest precision (98.59\%), ensuring highly accurate attack classification. Since our focus is on resource-constrained IoT devices in smart cities, we evaluated the model's training and inference times. While the training time of 122.5 seconds might seem long for on-device processing, training usually happens offline in real-world applications, so this factor doesn't affect our model's performance. More importantly, the inference time of 4 seconds is significantly faster than other deep learning models, even though we're not aiming for real-time processing.

\begin{table}[t!]
\centering
\renewcommand{\arraystretch}{1.4}
\caption{Comparing Accuracy: Proposed CNN-BiLSTM Model and Recent Works with Hybrid CNN and LSTM Architecture for Binary Classification in the UNSW-NB15 Dataset}
\label{tab:comparaison}
\begin{tabular}{>{\centering\arraybackslash}p{1cm} l p{1.8cm} l p{1.2cm}}
\toprule
\textbf{Ref} & \textbf{Year} & \textbf{Dataset} & \textbf{Algorithm} & \textbf{Accuracy}\\ 
\bottomrule
\textbf{\cite{9649612}} & 2021 & CIC-IDS2017 & BiLSTM & 97.72\% \\
\textbf{\cite{abdelkhalek2023addressing}} & 2023 & NSL-KDD & CNN-BiLSTM & 99.9\%\\
\multirowcell{3}{\textbf{\cite{app132111629}}} & \multirowcell{3}{2023} & NSL-KDD &  & 99.4\% \\
 & & UNSW-NB15 & \makecell[l]{CNN-BiLSTM} & 82.3\% \\
 & & CIC-IDS2017 & & 99.53\% \\
\textbf{\cite{ALTUNAY2023101322}} & 2023 & UNSW-NB15 & CNN-LSTM & 93.21\%\\
\textbf{Proposed model} & 2024 & UNSW-NB15 & CNN-BiLSTM & 97.28\%  \\ \bottomrule
\end{tabular}
\vspace{-4mm}
\end{table}

In Table \ref{tab:comparaison}, we summarize and compare the accuracy of our proposed CNN-BiLSTM model with recent works that adopt the same CNN-LSTM architecture. We observe that this architecture has already been used to improve the accuracy of IDS, not only on the UNSW-NB15 dataset but also on other traditional datasets in the field. In contrast, this work focuses on designing a lightweight hybrid model for resource-constrained devices using the BiLSTM. Regarding performance, we achieved a higher accuracy compared to works based on the UNSW-NB15 dataset. However, some works using the same architecture have achieved higher accuracy on the NSL-KDD dataset. This might be attributed to the larger number of features in UNSW-NB15 compared to NSL-KDD, potentially requiring more sophisticated feature selection for optimal performance.

\begin{table*}[]
    \centering
    \vspace{3mm}
    \renewcommand{\arraystretch}{1.3}
    \caption{Performance Comparison of CNN-BiLSTM and Traditional ML Models for Multiclassification in UNSW-NB15}
    \label{tab:multiclass}
    \begin{tabular}{lcccccc}
    \toprule
 & \textbf{Accuracy} & \textbf{Precision} & \textbf{Recall} & \textbf{F1-Score} & \textbf{Time to Train (s)} & \textbf{Time to Predict (s)} \\ \bottomrule
\textbf{Logistic} & 91.83\% & 94.31\% & 92.74\% & 93.53\% & 2.20 & 0.10 \\
\textbf{KNN} & 93.06\% & 95.06\% & 93.80\% & 94.43\% & 0.30 & 10.10 \\
\textbf{Decision Tree} & 84.10\% & 89.89\% & 86.28\% & 88.09\% & 0.20 & 0.10 \\
\textbf{GBC} & 75.47\% & 85.16\% & 79.11\% & 82.09\% & 55.90 & 0.10 \\
\textbf{MLP} & 96.74\% & 97.45\% & 97.00\% & 97.22\% & 122.80 & 0.10 \\
\textbf{LSTM} & 96.11\% & 97.02\% & 96.43\% & 96.72\% & 90.80 & 15.10 \\
\textbf{Proposed Model} & 96.91\% & 97.54\% & 97.13\% & 97.33\% & 1220.20 & 3.20 \\ \bottomrule
\end{tabular}
\vspace{-3mm}
\end{table*}

Table \ref{tab:multiclass} shows the results of the multiclassification of our CNN-BiLSTM model compared to the traditional model. To obtain these results, we made some adjustments to our model compared to the binary classification while keeping the same architecture. The adjustments involved the hyperparameters of the model, including the number of epochs, batch size, and learning rate. These changes affected the training and prediction times.
The results of the multiclassification were lower than those obtained in the binary classification due to the complexity of the task, which involves classifying 10 attacks using the proposed model. However, the obtained results demonstrate that our model outperforms other models in terms of accuracy, precision, recall, and F1-score, which are respectively 96.91\%, 97.54\%, 97.13\%, and 97.33\%.
Although the proposed model requires more time in the training phase due to adjustments in the number of epochs and batch size, it is always performed offline before deploying the model on resource-constrained IoT devices. Once deployed, these devices will act as IDS, and even with the increased training time and resource consumption, our model still meets the requirements of these devices in terms of resource usage due to the relatively lower time and resource needed for prediction.

\begin{figure}[t!]
	\centering
	\includegraphics[width=0.97\linewidth]{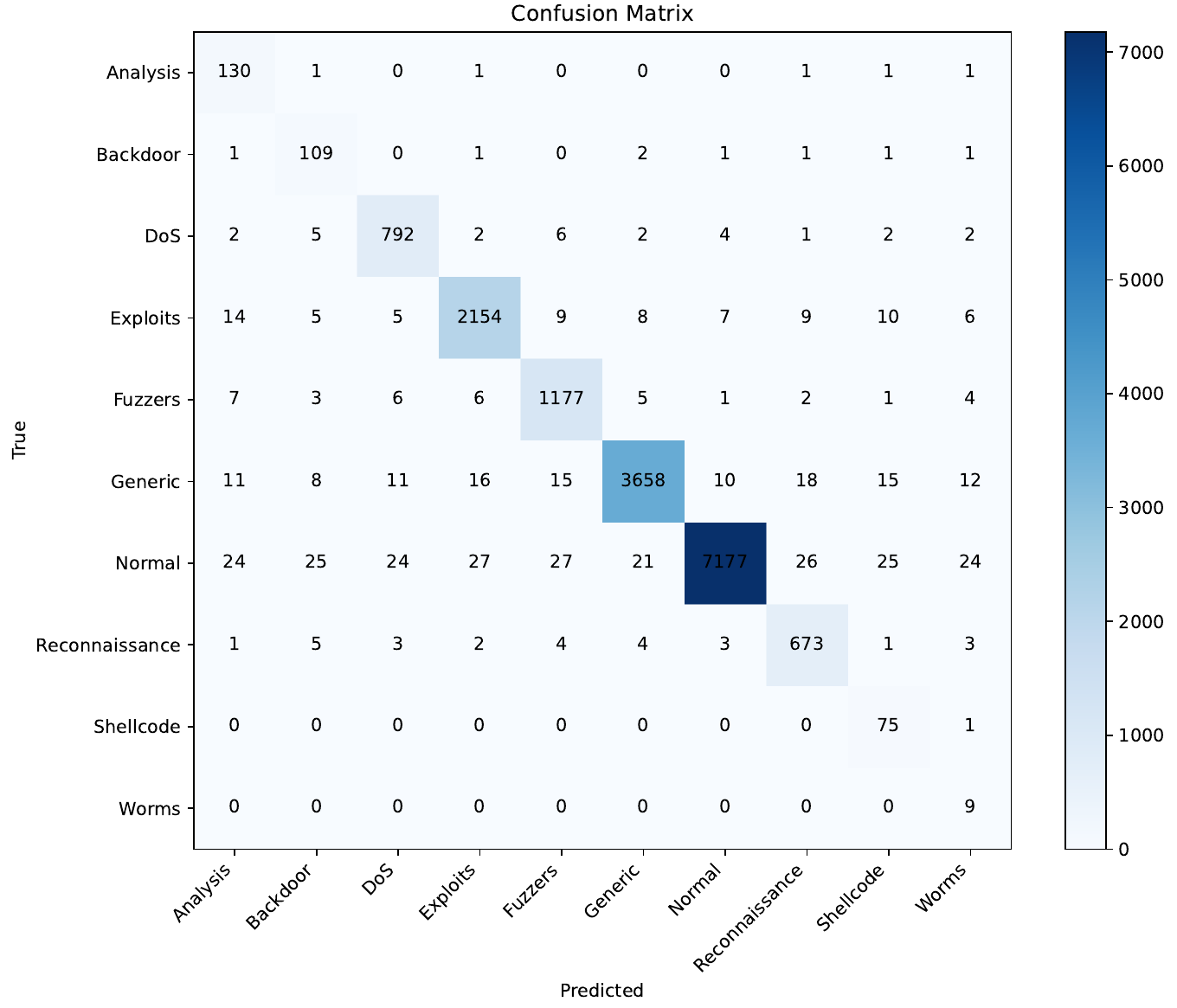}
	\caption{CNN-BiLSTM confusion matrix for multiclassification of attacks.}
	\label{matrix_multi}
 \vspace{-2mm}
\end{figure}

Figure \ref{matrix_multi} presents the confusion matrix for our CNN-BiLSTM multiclassification model on the UNSW-NB15 dataset, allowing us to evaluate each attack class individually. The dataset contains 10 attack classes, as shown in the x and y labels. The x-axis represents the ground truth (true labels), while the y-axis shows the predicted classes by our model. The diagonal, running from top-left to bottom-right, reflects the model's effectiveness in classifying each attack type. Darker squares indicate higher accuracy for that specific class. However, for minority classes like "Analysis," "Shellcode," and "Worms," the square colors might be less reliable due to the limited number of samples in the test data. By using a weighted loss function, our model prioritizes achieving better accuracy on these minority classes. This is evident in the correct detection of all 9 worms and the classification of 75 out of 76 Shellcode attacks with only one false negative. Conversely, the normal traffic class has the darkest square due to its high accuracy and abundance of samples. Overall, the confusion matrix serves as a visual representation of the model's accuracy. Generally, a darker diagonal signifies higher overall performance.

\section{Conclusion}
In this study, we developed a lightweight CNN-BiLSTM model tailored for resource-constrained IoT devices, addressing a critical gap in the application of complex machine learning models for IDS. Traditional models, while effective, are often unsuitable for such devices due to their computational demands. Our model not only considers model complexity but also achieves high accuracy, with binary and multiclass classification accuracies of 97.28\% and 96.91\%, respectively, on the UNSW-NB15 dataset. When compared to recent hybrid CNN-LSTM architectures, our model demonstrates superior performance on the UNSW-NB15 dataset, despite some models achieving up to 99\% accuracy on the NSL-KDD dataset. Moving forward, we aim to refine our approach by exploring Bayesian Optimization for hyperparameter tuning to enhance the performance of simpler models, thereby advancing research in IDS for IoT environments.

\vspace{-3mm}
\bibliographystyle{ieeetr}
\bibliography{library}

\end{document}